\documentclass{ws-ijmpa}

\begin{document}

\title{THERMAL CASIMIR FORCE\\
BETWEEN MAGNETIC MATERIALS}

\author{G. L. KLIMCHITSKAYA,$^{1,2}$ B. GEYER$^{2}$
and V. M. MOSTEPANENKO$^{3,2}$}

\address{
${}^1${North-West Technical University,
Millionnaya St. 5, St.Petersburg,
191065, Russia} \\
${}^2$Institute for Theoretical
Physics, Leipzig University, Postfach 100920,
D-04009, Leipzig, Germany \\
${}^3${Noncommercial Partnership ``Scientific Instruments'',
Tverskaya St. 11, Moscow,
103905, Russia}
}
\maketitle
\begin{abstract}
We investigate the Casimir pressure between two parallel plates
made of magnetic materials at nonzero temperature. It is shown
that for real magnetodielectric materials only the magnetic
properties of ferromagnets can influence the Casimir pressure.
This influence is accomplished through the contribution of the
zero-frequency term of the Lifshitz formula. The possibility
of the Casimir repulsion through the vacuum gap is analyzed
depending on the model used for the description of the dielectric
properties of the metal plates.
\end{abstract}



\section{Introduction}

The Casimir effect which results in a force acting between two
parallel electrically neutral material plates separated with a gap
of width $a$ finds many prospective applications ranging from
fundamental physics to nanotechnology.\cite{1}
For real dissimilar material plates described by the dielectric
permittivities $\varepsilon^{(1,2)}(\omega)$ and magnetic
permeabilities $\mu^{(1,2)}(\omega)$ at temperature $T$
in thermal equilibrium the
generalized Lifshitz formula for the Casimir (van der Waals)
pressure takes the form\cite{2}\cdash\cite{5}
\begin{eqnarray}
P(a,T)&=&-\frac{k_BT}{\pi}
\sum_{l=0}^{\infty}{\vphantom{\sum}}^{\prime}
\int_{0}^{\infty}\!\! q_lk_{\bot}dk_{\bot}\left\{
\left[\frac{e^{2aq_l}}{r_{\rm TM}^{(1)}(i\xi_l,k_{\bot})
r_{\rm TM}^{(2)}(i\xi_l,k_{\bot})}-1\right]^{-1}\right.
\nonumber \\
&&\left.
+\left[\frac{e^{2aq_l}}{r_{\rm TE}^{(1)}(i\xi_l,k_{\bot})
r_{\rm TE}^{(2)}(i\xi_l,k_{\bot})}-1\right]^{-1}\right\}.
\label{eq1}
\end{eqnarray}
\noindent
Here, $k_B$ is the Boltzmann constant, prime adds a multiple
one half to the term with $l=0$, $k_{\bot}$ is the projection
of the wave vector on the plane of the plates, and
$\xi_l=2\pi k_BTl/\hbar$ with $l=0,\,1,\,2,\,\ldots$ are
the Matsubara frequencies. The reflection coefficients for
the transverse magnetic and transverse electric polarizatons
of the electromagnetic field are given by
\begin{equation}
r_{\rm TM}^{(n)}(i\xi_l,k_{\bot})=
\frac{\varepsilon_l^{(n)}q_l-k_l^{(n)}}{\varepsilon_l^{(n)}q_l
+k_l^{(n)}}, \qquad
r_{\rm TM}^{(n)}(i\xi_l,k_{\bot})=
\frac{\mu_l^{(n)}q_l-k_l^{(n)}}{\mu_l^{(n)}q_l+k_l^{(n)}},
\label{eq2}
\end{equation}
\noindent
where $\varepsilon_l^{(n)}\equiv\varepsilon^{(n)}(i\xi_l)$,
$\mu_l^{(n)}\equiv\mu^{(n)}(i\xi_l)$, the index $n=1,\,2$
numerates the plates and
\begin{equation}
q_l^2=k_{\bot}^2+\frac{\xi_l^2}{c^2}, \qquad
{k_l^{(n)}}^2=k_{\bot}^2+
\varepsilon_l^{(n)}\mu_l^{(n)}\frac{\xi_l^2}{c^2}.
\label{eq3}
\end{equation}

As was noticed long ago,\cite{6} ``In the majority of cases,
the contribution to the van der Waals interaction due to the
magnetic properties of real materials is extremely small.''
A large contribution (including the Casimir repulsion through
a vacuum gap for some range of parameters) was found\cite{7}
using the approximation of frequency-independent
$\varepsilon$ and $\mu$. Later, however, it was shown\cite{8}
that for real materials $\mu$ is nearly equal to unity in the
range of frequencies which gives a major contribution to the
Casimir pressure. This problem was reconsidered\cite{4}
at both zero and nonzero temperature for one metallic and
one magnetodielectric plate using the description of a metal by
means of the Drude model and of magnetodielectric by a
simplified model of the Drude-Lorentz type. At $T\neq 0$ the
Casimir force was found to be always attractive.

In this paper we investigate the thermal Casimir pressure
between plates made of ferromagnetic metal, ferromagnetic
dielectric and nonmagnetic metal taking into account
realistic dependences of $\varepsilon$ and $\mu$ on the
frequency and using different approaches to the theory of
the thermal Casimir force suggested in the literature.
We demonstrate how the use of different approaches
influences the Casimir pressure and find when the
Casimir repulsion through a vacuum gap is feasible.
In Sec.~2 we provide a brief review of magnetic properties.
Sec.~3 deals with ferromagnetic metals and Sec.~4 with
ferromagnetic dielectrics. In Sec.~5 we consider the
behavior of the Casimir pressure in the vicinity of Curie
temperature. Sec.~6 contains our conclusions.

\section{Review of magnetic properties}

The magnetic permeability along the imaginary frequency axis
is represented in the form
\begin{equation}
\mu(i\xi)=1+4\pi\chi(i\xi),
\label{eq4}
\end{equation}
\noindent
where $\chi(i\xi)$ is the magnetic susceptibility.
The magnitude of $\chi(i\xi)$ decreases monotonously when
$\xi$ increases. All materials possess diamagnetic polarization
for which\cite{9}\cdash\cite{11} $\chi(0)<0$, $\mu(0)<1$
and $|\mu(0)-1|\sim 10^{-5}$. Diamagnets (such materials as,
for instance, Au, Si, Cu and Ag) do not possess any other
type of magnetic polarization. For them one can put
$\mu_l=1$, $l=0,\,1,\,2,\,\ldots$ in computations using
(\ref{eq1}) so that magnetic properties of diamagnets do not
influence the Casimir force.

Some materials also possess paramagnetic polarization (in a
broad sense) which is larger in magnitude than the diamagnetic
one and leads to\cite{11}$\chi(0)>0$, $\mu(0)>1$.
Paramagnets (in a narrow sense) are materials with $\mu(0)>1$
if the interaction of magnetic moments of their constituent
particles is neglibibly small. Paramagnets may consist of
microparticles which are paramagnetic magnetizable but have no
intrinsic magnetic moment (the Van Vleck polarization
paramagnetism\cite{12}) and of microparticles possessing
a permanent magnetic moment (the orientational
paramagnetism\cite{9}\cdash\cite{12}). For all paramagnets in a
narrow sense  it is true that
$\chi(0)<10^{-4}$ and one can put $\mu_l=1$ for all $l$.
This conclusion is unchanged for all paramagnets in a broad sense
(with the single exception of ferromagnets) because $\chi(0)$
remains as small as mentioned above and takes only a
slightly larger
values in the vicinity of $T=0$
even at temperatures below
the critical temperature $T_{\rm cr}$ of the magnetic phase
transitions \cite{9}\cdash\cite{11,13}\cdash\cite{15}
(for different materials $T_{\rm cr}$ varies from a few K to more
than thousand K).

For the subdivision of paramagnetic materials called ferromagnets
it is true that
$\mu(0)\gg 1$ at $T<T_{\rm cr}$ (in this case $T_{\rm cr}$
is referred to as the Curie temperature, $T_{\rm cr}\equiv T_C$).
There is a lot of ferromagnetic materials, both metals and
dielectrics.\cite{16} The rate of decrease of $\mu(i\xi)$ for
ferromagnets depends on their electric resistance. Thus, for
ferromagnetic metals and dielectrics $\mu(i\xi)$ becomes
approximately equal to unity for $\xi$ above $10^4$
and $10^9\,$Hz, respectively. Keeping in mind that the first
Matsubara frequency $\xi_1\sim 10^{14}\,$Hz at $T=300\,$K
we arrive at the conclusion that ferromagnets can affect the
Casimir force between macroscopic bodies only through the
contribution of the zero-frequency term of the Lifshitz formula
(\ref{eq1}). In all terms of this formula with $l\geq 1$ one can
put $\mu_l=1$. Note that below we do not consider so-called
hard ferromagnetic materials possessing a spontaneous
magnetization because the magnetic interaction between
the plates made of such materials far exceeds any conceivable
Casimir force. The subject of our interset is the soft
ferromagnetic materials which do not possess a spontaneous
magnetization. It is well known that the magnetic permeability
of ferromagnets depends on the applied magnetic
field.\cite{9}\cdash\cite{11} Since in the Casimir interaction
the mean applied field is equal to zero, below we consider
the so-called {\it initial} (zero field) permeability, i.e.,
$\mu=\mu(\mbox{\boldmath$H$}=0)$.

\section{Ferromagnetic metals}

First we consider the case when both Casimir plates are made
of common ferromagnetic metal Co, Cd, Fe or Ni.
The dielectric properties of a metal are described by the
Drude\cite{17,18} or the plasma\cite{19,20} model approaches,
i.e., using the dielectric functions of the form
\begin{equation}
\varepsilon_D(i\xi)=1+\frac{\omega_p^2}{\xi(\xi+\gamma)},
\qquad
\varepsilon_D(i\xi)=1+\frac{\omega_p^2}{\xi^2},
\label{eq5}
\end{equation}
\noindent
where $\omega_p$ is the plasma frequency, $\gamma$ is the
relaxation parameter. By considering different models
proposed in the literature we aim to determine whether or not
the magnetic properties influence the magnitude of the
Casimir pressure and is it possible to experimentally
distinguish between alternative theoretical predictions.
For all $l\geq 1$ we put $\mu_l=1$.

For two similar plates Eq.~(\ref{eq2}) leads to the following
reflection coefficients at $\xi=0$ if the Drude and plasma
models are used
\begin{eqnarray}
&&
r_{{\rm TM},D}(0,k_{\bot})=r_{{\rm TM},p}(0,k_{\bot})=1,
\quad
r_{{\rm TE},D}(0,k_{\bot})=
\frac{\mu(0)-1}{\mu(0)+1}\equiv r_{\mu},
\nonumber \\
&&
r_{{\rm TE},p}(0,k_{\bot})=\frac{\mu(0)ck_{\bot}-
[c^2k_{\bot}^2+\mu(0)\omega_p^2]^{1/2}}{\mu(0)ck_{\bot}+
[c^2k_{\bot}^2+\mu(0)\omega_p^2]^{1/2}}.
\label{eq6}
\end{eqnarray}
\noindent
In the limiting case of large separations (high $T$) only the
zero-frequency term in Eq.~(\ref{eq1}) contributes
to the Casimir pressure and all calculations can be performed
analytically. When the Drude model is used, the result is
\begin{equation}
P_{D}(a,T)=-\frac{k_BT}{8\pi a^3}\left[\zeta(3)+
\mbox{Li}_3(r_{\mu}^2)\right],
\label{eq7}
\end{equation}
\noindent
where $\zeta(z)$ is the Riemann zeta function and
$\mbox{Li}_n(z)$ is polylogarithm function.
Under the conditions $\mu(0)\gg 1$ (valid for ferromagnetic
metals) and $\mu(0)=1$ (valid for nonmagnetic
metals) Eq.~(\ref{eq7}) leads to
\begin{equation}
P_{D,{\rm fm}}(a,T)=-\frac{k_BT}{4\pi a^3}\zeta(3),
\qquad
P_{D,{\rm nm}}(a,T)=-\frac{k_BT}{8\pi a^3}\zeta(3),
\label{eq8}
\end{equation}
\noindent
respectively. As can be seen from Eq.~(\ref{eq8}), if the Drude
model is used, the account of magnetic properties of ferromagnetic
metals doubles the magnitude of the Casimir pressure at large
separations. If, however, the plasma model is used at large
separations under the condition
$\sqrt{\mu(0)}\delta_0/a\ll 1$, Eq.~(\ref{eq1}) results in
\begin{equation}
P_{p,{\rm fm}}(a,T)=-\frac{k_BT}{4\pi a^3}\zeta(3)
\left[1-3\sqrt{\mu(0)}\frac{\delta_0}{a}\right],
\label{eq9}
\end{equation}
\noindent
where $\delta_0=c/\omega_p$ is the skin depth.
The same expression, but with $\mu(0)=1$, is obtained for
nonmagnetic metals described by the plasma model.
At $T=300\,$K Eqs.(\ref{eq7})--(\ref{eq9}) are
applicable for $a>6\,\mu$m.

\begin{figure}[b]
\vspace*{-13.7cm}
\hspace*{-5.1cm}
\psfig{file=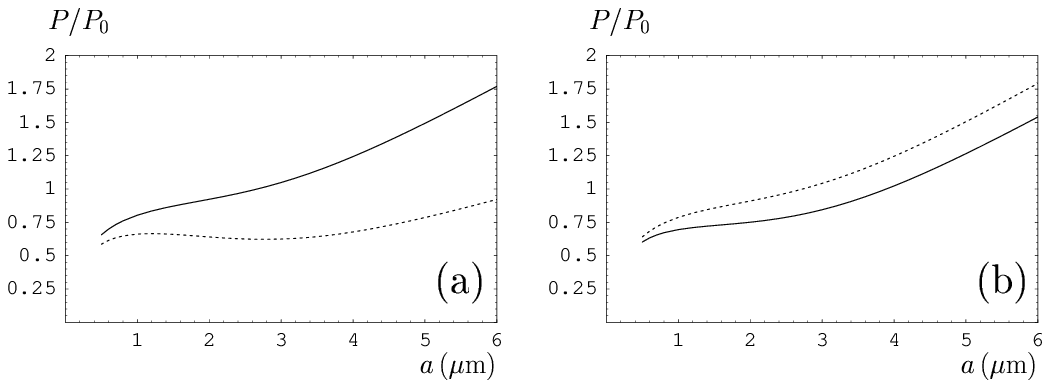,width=8.5in}
\vspace*{-13.5cm}
\caption{The relative Casimir pressure as a function of separation
in the configuration of two parallel Co plates with
inclusion of
magnetic properties (the solid lines) and with magnetic
properties neglected (the dashed lines). Computations are
performed with the dielectric permittivity  (a) of the Drude model and
(b) of the plasma model.}
\label{aba:fig1}
\end{figure}

{}From an experimental point of view the most interesting region
is from $a=0.5\,\mu$m to $a=1\,\mu$m (for $a<0.5\,\mu$m the
contribution of the zero-frequency term and, thus, of magnetic
properties is not large enough).
We have performed numerical computations of the Casimir pressure,
Eq.~(\ref{eq1}), in the region from 0.5 to $6\,\mu$m for Co
with parameters\cite{21,22}
$\omega_{p,{\rm Co}}=3.97\,$eV, $\gamma_{\rm Co}=0.036\,$eV
and $\mu_{\rm Co}(0)=70$. In Fig.~1 we plot the ratio of the
Casimir pressure $P$ between two Co plates at $T=300\,$K
to $P_0=-\pi^2\hbar c/(240a^4)$ computed using (a) the Drude
model and (b) the plasma model. The solid lines take into account
the magnetic properties and the dashed lines are computed with
magnetic properties neglected. Note that the solid line
in Fig.~1(a) is almost coincident with the dashed line in
Fig.~1(b). At small separations $a<1\,\mu$m, the difference
between the dashed line in Fig.~1(a) and the solid line
in Fig.~1(b) is also not observable in the limits of the
experimental precision. Thus the experiments on an indirect
measurement of the Casimir pressure by means of a micromechanical
ocsillator\cite{23} at separations of about 500--600\,nm
can allow us to choose one of the following situations.

1. The experimental data are in favour of the solid line
in Fig.~1(a) and the dashed line in Fig.~1(b). This means
that either the magnetic properties affect the Casimir
pressure and metals should be described by the Drude model
or the magnetic properties do not affect the Casimir
pressure and metals should be described by the plasma model.

2. The experimental data are in favour of the dashed line
in Fig.~1(a) and the solid line in Fig.~1(b). In this
case either the magnetic properties affect the Casimir
pressure and metals should be described by the plasma model
or the magnetic properties do not affect the Casimir
pressure and metals should be described by the Drude model.

Now let one plate be made of a ferromagnetic metal ($n=1$)
and the other of a nonmagnetic metal ($n=2$). Here, in the
limit of large separations one obtains
\begin{equation}
P_D(a,T)=-\frac{k_BT}{8\pi a^3}\zeta(3)
\label{eq10}
\end{equation}
\noindent
if the Drude model is used. For the plasma model under
conditions
$\sqrt{\mu(0)}\delta_{01}/a\ll 1$ and
$\delta_{02}/a\ll 1$ it follows
\begin{equation}
P_{p}(a,T)=-\frac{k_BT}{4\pi a^3}\zeta(3)
\left[1-
\frac{3(\sqrt{\mu(0)}\delta_{01}+\delta_{02})}{2a}\right].
\label{eq11}
\end{equation}
\noindent
Note than when the Drude model is used
$r_{\rm TE}^{(2)}(0,k_{\bot})=0$ and, thus,
the magnetic properties of a ferromagnetic plate entering only
through $r_{\rm TE}^{(1)}(0,k_{\bot})$ do not influence
the result.
The results of the numerical computations for the Co plate interacting
with the Au plate
($\omega_{p,{\rm Au}}=9.0\,$eV, $\gamma_{{\rm Au}}=0.035\,$eV)
in the case when the plasma model is used are shown in Fig.~2(a).
\begin{figure}[t]
\vspace*{-13.7cm}
\hspace*{-5.1cm}
\psfig{file=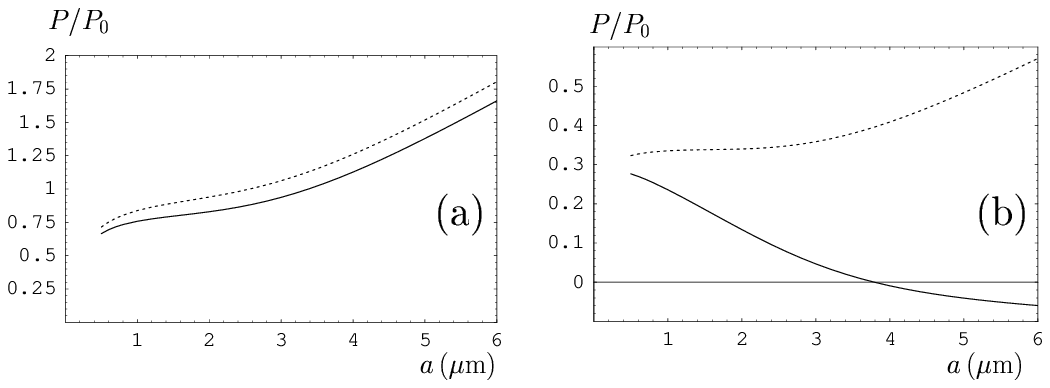,width=8.5in}
\vspace*{-13.5cm}
\caption{The relative Casimir pressure as a function of separation
in the configuration of two parallel plates, one made of Au and the other
of (a) Co and (b) ferromagnetic dielectric with inclusion of
magnetic properties (the solid lines) and with magnetic
properties neglected (the dashed lines). Computations are
performed using the plasma model for the dielectric permittivity
of metals.}
\label{aba:fig2}
\end{figure}
It is seen that here the inclusion of the magnetic properties
(the solid line) decreases the magnitude of the Casimir
pressure. The influence of the magnetic properties is, however,
very moderate and can be observed only in the experiment
on measuring the difference Casimir pressure above a
patterned plate\cite{24} one section of which is made of
Co and the other  of Au. Such an experiment allows one to
choose between the two alternatives in each of the situations
described above. This will provide a complete
experimental answer to questions whether the magnetic
properties influence the Casimir force and what
dielectric model should be used in the Lifshitz theory
to describe real metals.

\section{Ferromagnetic dielectrics}

Ferromagnetic dielectrics are materials that, while
displaying physical properties characteristic of dielectrics,
show ferromagnetic behavior under the influence of an external
magnetic field.\cite{25} Such materials are widely used in
different magneto-optical devices. As an  example, we consider
a composite material of polystyrene with a
volume fraction $f$ of ferromagnetic metal nanoparticles in
the mixture. The permittivity of such a material can be
presented in the form\cite{26}
\begin{equation}
\varepsilon_{\rm fd}(i\xi)=\varepsilon_{\rm d}(i\xi)
\left(1+\frac{3f}{1-f}\right),
\label{eq12}
\end{equation}
\noindent
where $\varepsilon_{\rm d}$ is the permittivity of
polystyrene.\cite{27}

We have performed computations of the Casimir pressure for
two parallel plates one of which is made of ferromagnetic
dielectric [$f=0.25$, $\varepsilon_{\rm fd}(0)=5.12$,
$\mu(0)=25$] and the other of Au described by the plasma model.
Recall that if Au is described by the Drude model the
magnetic properties do not influence the Casimir pressure
as explained in Sec.~3. The computational results for
$P/P_0$ as a function of $a$ are presented in Fig.~2(b)
where the solid line takes the magnetic properties into
account and the dashed line neglects them.
As can be seen in Fig.~2(b), magnetic properties
have an important influence
 on the Casimir pressure and even lead to the
change of sign of the force (from attraction to repulsion).

This important conclusion can be confirmed analytically in the
limiting case of large $a$. If the metallic properties of Au
plate are described by the Drude model, one obtains
\begin{equation}
P_D(a,T)=-\frac{k_BT}{8\pi a^3}\mbox{Li}_3(r_{\varepsilon}),
\qquad
r_{\varepsilon}\equiv
\frac{\varepsilon_{\rm fd}-1}{\varepsilon_{\rm fd}+1}.
\label{eq13}
\end{equation}
\noindent
This does not depend on the magnetic properties. If, however,
the plasma model is used, then, under the condition
$\delta_{02}/a\ll 1$, one arrives at
\begin{equation}
P_p(a,T)=-\frac{k_BT}{8\pi a^3}\left[\mbox{Li}_3(r_{\varepsilon})
+\mbox{Li}_3(-r_{\mu})\left(1-
3\frac{\delta_{02}}{a}\right)\right].
\label{eq14}
\end{equation}
\noindent
The expression on the right-hand side of Eq.~(\ref{eq14}) is
positive and the respective Casimir force is repulsive if the
following condition is satisfied:
\begin{equation}
\mbox{Li}_3(r_{\varepsilon})<
\left|\mbox{Li}_3(-r_{\mu})\left(1-
3\frac{\delta_{02}}{a}\right)\right|.
\label{eq15}
\end{equation}
\noindent
This condition is easily satisfied for real materials.

\section{Vicinity of the Curie temperature}

At the Curie temperature $T_C$ specific for each material,
ferromagnets undergo a magnetic phase transition.\cite{14,16}
At higher temperature they lose ferromagnetic properties and
become paramagnets in the narrow sense. Thus, for Fe, Co, Ni
and Gd the Curie temperature is equal to 1043\,K, 1388\,K, 627\,K
and 293\,K, respectively.\cite{28} Here, we consider the
behavior of the Casimir pressure under the magnetic phase
thansition which occurs with the increase of $T$ in the
configuration of two similar plates made  of Gd. The Drude
parameters of Gd are equal to\cite{29}
$\omega_{p,{\rm Gd}}=9.1\,$eV, $\gamma_{{\rm Gd}}=0.58\,$eV.

Computations of the Casimir pressure between two parallel plates
made of Gd in the vicinity of the Curie temperature require respective
values of $\mu(0)$ for Gd at $T<T_C$ [at $T>T_C$,
$\mu_{\rm Gd}(0)=1$ to high accuracy].
In Fig.~3(a) the magnetic permeability of Gd is shown as a
function of temperature in the region from 280\,K to
300\,K on the basis of the experimental data.\cite{30}
\begin{figure}[t]
\vspace*{-13.7cm}
\hspace*{-5.1cm}
\psfig{file=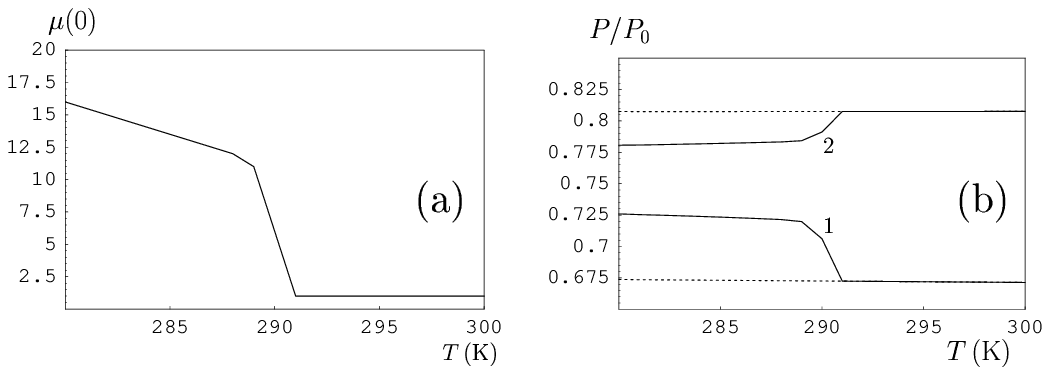,width=8.5in}
\vspace*{-13.5cm}
\caption{(a) The static magnetic permeability of Gd
at the magnetic phase transition as a function of temperature.
(b)
The relative Casimir pressure as a function of temperature
in the configuration of two parallel Gd plates at the separation
$a=0.5\,\mu$m. The solid and dashed lines include and
neglect the magnetic properties, respectively. The pairs
of lines marked 1 and 2 indicate the respective computational
results obtained using the Drude and the plasma models.
}
\label{aba:fig3}
\end{figure}
The Casimir pressure as a function of temperature was computed
at the separation $a=500\,$nm between the plates using
Eq.~(\ref{eq1}). The computational results obtained using
the Drude and the plasma models are shown in Fig.~3(b) by
the pairs of
lines 1 and 2, respectively. In each pair the solid line
takes into account the magnetic properties and the dashed line is
computed with these properties disregarded.
As can be seen from Fig.~3{b},
experiments on the magnetic phase transition can also
be used to determine the influence of magnetic properties on the
Casimir force and as a test for different models of the dielectric
properties of metals.

\section{Conclusions}

The investigation of the influence of magnetic properties on the
Casimir force performed above leads
to the following conclusions.
\begin{enumerate}
\item[1.]
Of all the real materials, only ferromagnets might affect
the Casimir force.

\item[2.]
At all feasible temperatures the possible influence of
ferromagnets on the Casimir force occurs solely through
the contribution of the zero-frequency term in the
Lifshitz formula.

\item[3.]
In the framework of the Lifshitz theory the Casimir
repulsion of two macroscopic bodies separated by a vacuum
gap arises for only the case when one body is made of
ferromagnetic dielectric and the other is metallic. In doing
so the metal is described by the plasma model.

\item[4.]
Modern experimental techniques present good
opportunities to check
whether the magnetic properties of the plate material influence
the Casimir force. Experiments with magnetic bodies allow
independent test of the plasma and Drude model approaches
to the description of the  dielectric properties of metals.

\end{enumerate}

\section*{Acknowledgments}

The authors are grateful to the
Deutsche Forschungsgemeinschaft
Grant No.~GE\,696/9--1
for partial financial support.


\end{document}